\documentstyle[twocolumn,aps,prl,epsf]{revtex}
\tighten
\draft

\def\ha{Hamiltonian }
\def\<{\langle}
\def\>{\rangle}
\def\sus{susceptibility }
\def\suss{susceptibilities }

\def\half{{\frac{1}{2}}}
\def\be{\begin{equation}}
\def\ee{\end{equation}}
\def\imchi{\chi^{\prime \prime}(\nu)}

\begin{document}
\preprint{cond-mat}
\title{Linear and Nonlinear Susceptibilities\\of a
Decoherent Two Level System}

\author{Gregory Levine}

\address{Department of Physics, Hofstra University, Hempstead, NY 11550\\and}
\address{Department of Physics, Brookhaven National Laboratory, Upton, NY 11973-5000}
\date{\today}
\maketitle
\begin{abstract}
The linear and nonlinear dynamical susceptibilities of a two level
system are calculated as it undergoes a transition to a decoherent
state.  Analogously to the Glover-Tinkham-Ferrell sum rule of
superconductivity, spectral weight in the linear susceptibility is
continuously transferred from a finite frequency resonance to nearly
zero frequency, corresponding to a broken symmetry in the
thermodynamic limit.  For this reason, the behavior of the present
model (the Mermin Model) differs significantly from the spin-boson
model.  The third order nonlinear susceptibility, corresponding to
two-photon absorption, has an unexpected non-monotonic behavior as a
function of the environmental coupling, reaching a maximum within the
decoherent phase of the model.  Both linear and nonlinear
susceptibilities may be expressed in a universal form.
\end{abstract}


\section{Introduction}
\label{intro}

A vast body of work has been devoted to understanding the transition
to decoherence in models of a two-level-system (TLS) coupled to an
infinite set of environmental degrees of freedom
\cite{leggett_rmp}. The central quantity in this transition is the
dynamical correlation function, or equivalently, the linear
susceptibility, $\imchi$, of the system degrees of freedom.  To our
knowledge, no studies of the nonlinear susceptibility have been
undertaken.  Our interest in the nonlinear response is twofold: 1)
experimental probes of Macroscopic Quantum Coherence (MQC) might
access the nonlinear regime and 2) the nonlinear susceptibility
appears to bear a different relationship to the coherence of the TLS
than the linear susceptibility.  A measure of coherence of the system
is the quantity of spectral weight within the resonance peak of
$\imchi$ associated with the TLS---this is simply proportional to the
probability of absorbing a {\sl single} photon.  Absorption of a
photon at a precise energy is a property uniquely associated with the
underlying coherence of the quantum mechanical system.  As the
coupling to the environment is increased and the system increasingly
localizes in one state, the probability of single photon absorption
diminishes.  In this paper we address whether {\sl two photon}
absorption is qualitatively similar---is it similarly linked to the
underlying coherence of the system?  We are thus led to the study of
nonlinear susceptibilities of a TLS as the system is tuned through a
decoherence transition.

N. D. Mermin investigated a model obtained from the well known
spin-boson Hamiltonian by keeping only the lowest two levels of each
harmonic oscillator comprising the environmental bath
\cite{mermin91}. The \ha therefore contains a ``system'' spin-1/2,
represented by the Pauli matrices $\bf \sigma$, coupled to a set of
$N$ ``environment'' spin-1/2 degrees of freedom, $\{{\bf s}_j\}$. The
resulting Hamiltonian
\be
\label{ham}
H=\frac{t}{2} \sigma_x + \frac{\lambda}{4N} \sigma_z
\sum_{j=1}^{N}{(s_j^+ + s_j^-)} + \frac{\omega}{2N}
\sum_{j=1}^{N}{s_j^z} \ee (henceforth referred to as the Mermin Model)
can be solved to demonstrate the close correspondence between
decoherence and a second order phase transition.  The transition in
the Mermin model also bears some similarity to the decoherence transition
in the Ohmic case of the spin-boson Hamiltonian
\cite{hanggi98}.  Owing to
its finite Hilbert space, exact dynamical susceptibilities (linear and
nonlinear) can be calculated and followed through the transition. The
finite dimensionality may also be realistic in some physical
settings and exhibits the expected corrections to an ``infinite
environment'' thermodynamic phase transition. In this paper, we will
present such calculations and a careful study of the Mermin Model at
the linear level, demonstrating an analogy to the superconducting
transition in the ``dirty'' limit.  We will then go on to the
calculation and interpretation of the third order, nonlinear
\sus corresponding to two photon absorption.

To briefly summarize our findings: The transition to decoherence is
induced by strong coupling to environmental degrees of freedom (large
$\lambda$) or a small system energy scale (small $t$).  Below a
critical coupling, $\lambda_c$, all spectral weight of the dynamical
\sus
\be 
\label{imchi}
\imchi \equiv \mbox{Im} \,\,\frac{i}{4}\int_{-\infty}^{\infty}{dt
\<0|[\sigma(t),\sigma(0)]|0\> \theta (t) e^{i\nu t}} \ee lies in the
principal resonance of the TLS at an energy $t$.  When the coupling is
increased above the critical coupling, a new exponentially small
energy scale $O(te^{-N/2})$ emerges, associated with a broken symmetry,
$\<\sigma\>\neq 0$, in the thermodynamic limit $N \rightarrow
\infty$.  This feature corresponds to tunneling modified by a
Franck-Condon type overlap factor.  As the coupling is increased,
spectral weight is continuously shifted to the ``near-zero'' frequency
channel.  The weight of the delta-function, $\delta(0^+)$, is simply
the order parameter, $|\<0|\sigma|0\>|^2$ (or, more exactly,
$|\<1|\sigma|0\>|^2$, for large but finite $N$). However, the
dynamical \sus obeys a sum rule implying that an incompletely formed
broken symmetry state leaves some spectral weight at the position of
the principal resonance, set by $t$.

It is in this respect that the Mermin model differs from the ohmic
case of the spin-boson Hamiltonian (SBH).  In the SBH, the Toulouse
limit (at $\alpha=1/2$), corresponding to complete inelastic
broadening of the resonance, is a precursor to localization of the
system spin (at $\alpha=1$) \cite{stockburger98,lesage96}.  In the
Mermin model, the quantum resonance of the system spin---although
damped---remains intact through the decoherence transition. This
behavior is analogous to the Glover-Tinkham-Ferrell \cite{gtf} sum
rule for the dynamical conductivity in superconductors.  Spectral
weight falling in the gap region, proportional to the superfluid
density (or, equivalently, the order parameter) is redistributed to a
delta-function at zero frequency. In the case of the transition to
decoherence, finite but supercritical coupling, $\lambda > \lambda_c$,
is analogous to the transition to superconductivity in the dirty
limit where the superconductor can absorb energy above the gap.

In contrast, the nonlinear response has a more complicated behavior.
Two photon absorption (TPA) requires the presence of ancillary levels
as intermediate states and is therefore enhanced when there is some
degree of ``inelastic'' broadening of the primary resonance.  At the
two extremes, TPA is zero when $\kappa > \kappa_c$ and the environment
is decoupled from the system---but it is also zero when $\kappa
\rightarrow 0$ and the system is completely decoherent.

We now turn to a review of the analytical results obtained from the
Mermin model.

\section{The Mermin Model}
\label{mermin}

In the absence of coupling to the environment, the two level system in
(\ref{ham}) possesses a ground state in a coherent superposition of spin
up and spin down (denoted $|+\>-|-\>$) and will display a sharp
resonance in $\imchi$ at $\nu = t$.  When environmental coupling is
included, there are two effects commonly associated with decoherence.
First, the coherence features should shift to lower frequency
(Franck-Condon overlap) and secondly, the
peak should broaden from ``inelastic'' energy exchange with the
environment.  These two effects are nicely demonstrated in the Mermin
Model.
In (\ref{ham}), the environment spins may be summed to one big
$O(N)$ spin, $\bf S$, and the \ha becomes
\be
\label{ham2}
H=\frac{t}{2} \sigma_x +  \frac{\lambda}{2N} \sigma_z S_x + \frac{\omega}{2N}
S_z \ee
As Mermin points out, the advantage of the \ha (\ref{ham2}) is that in
the limit $N \rightarrow \infty$, the environment spins may be
replaced in the \ha by the $x$ and $z$ components of a classical spin
angular momentum: $\frac{\lambda}{N} \sigma_z S_x
\rightarrow \half \lambda \sigma_z \sin{\theta}$ and $\frac{\omega}{N} S_z
\rightarrow -\half \omega \cos{\theta}$.  The resulting Hamiltonian
\be
\label{ham3}
H = \frac{t}{2}\sigma_x + \frac{\lambda}{4} \sigma_z \sin{\theta} -
\frac{\omega}{4} \cos{\theta} \ee 
may be diagonalized and its ground
state eigenvalue, $E_0(\theta)$, given by
\be
E_0(\theta) = -\half \sqrt{t^2 + \frac{\lambda^2}{4} \sin^2{\theta}}
- \frac{\omega}{4} \cos{\theta}
\ee
minimized with respect to $\theta$.

\begin{figure}[tb] 
\epsfxsize=3.3in 
%
\centerline{\epsfbox{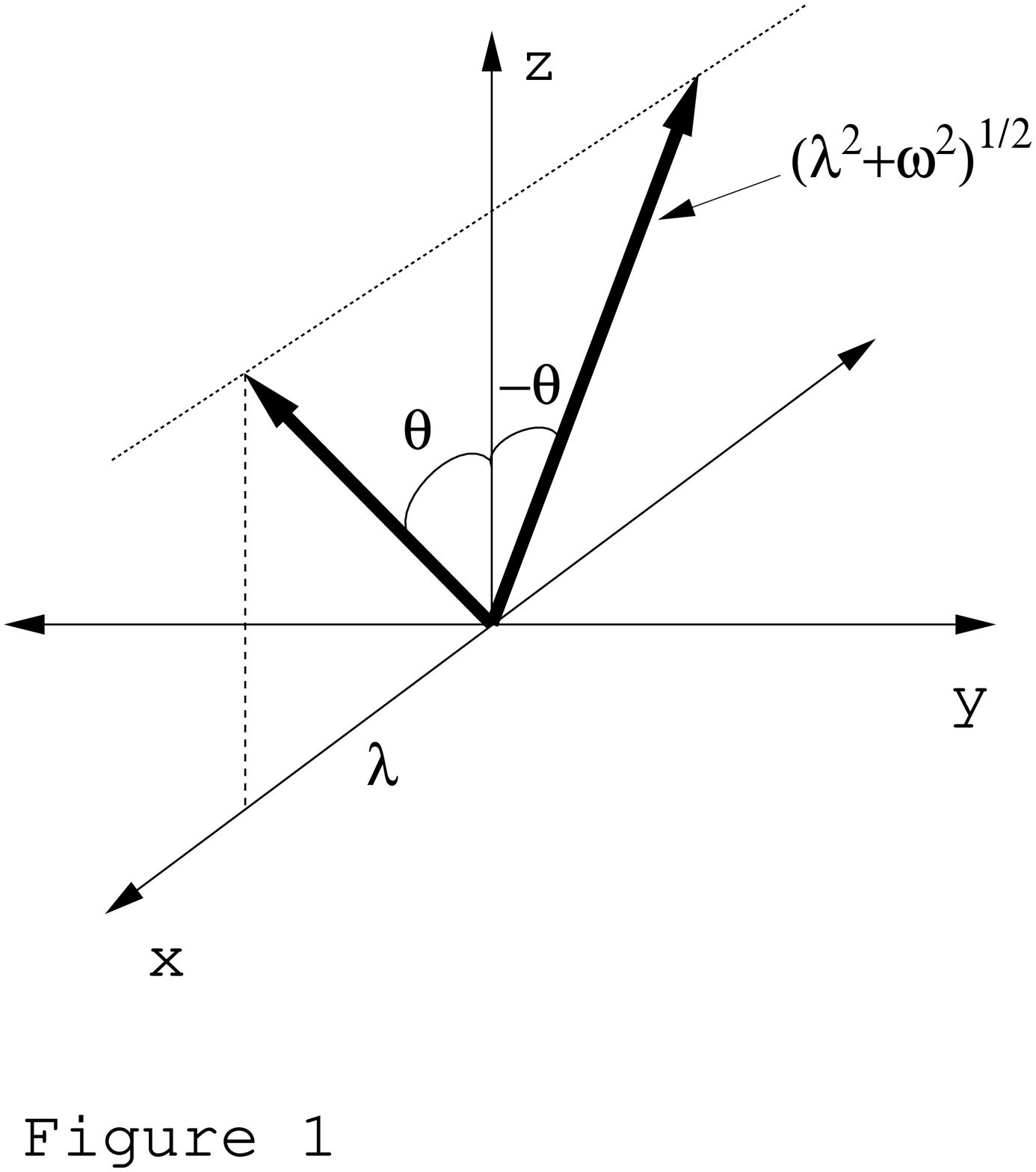}} 
\vskip 0.3truein \protect\caption{The classical ($N \rightarrow
\infty$) minimum energy configurations of the environment (solid black
arrows) in the strongly coupled phase, $\kappa<\kappa_c$.  ($\kappa
\equiv 2\omega t/\lambda^2$; $\kappa_c \equiv 1$).}
\end{figure} 

The critical behavior of the model may now be seen by examining the
ground state energy $E_0$, as function of $t$. $E_0$ bifurcates at a
finite value of $t$ going from a singlet, non-degenerate root for
$t>\lambda^2/2\omega$ to a doubly degenerate set for
$t<\lambda^2/2\omega$.
\begin{eqnarray}
\label{crit_point}
\theta_0 = 0~~~~~~~~~~~~~~~~~~~~~~~~~~~ &  t>\lambda^2/2\omega  \\
\sin{\theta_0} = \pm \sqrt{\frac{1-4t^2\omega^2/\lambda^4}
{1+\omega^2/\lambda^2}}~~~~~~ &  t<\lambda^2/2\omega 
\end{eqnarray}
In the former case, the environment is decoupled from the system and
always points along the $z$-axis to minimize its ``Zeeman'' energy
corresponding to the last term of (\ref{ham2}).  In the latter case, the
environment and the system spins are frozen in two distinct
orientations with degenerate energies.  The ground state wave function
is still in a superposition $\alpha|+\>-\beta|-\>$ but it is no longer an
evenly weighted one ($\alpha=\beta$); rather, the two roots correspond
to the system predominantly in  $|+\>$ or $|-\>$.

However, for finite (even very large) $N$, the system must be in an
evenly weighted superposition---independent of the value of $t$.  For
$t \rightarrow 0$, it is possible to estimate the tunnel splitting
between the ground and first excited states.  Following
\cite{mermin91}, we consider the adiabatic environment (or small $t$)
limit in which the environment instantaneously adjusts to the
localization of the system in a given state and thus ``points'' at the
angle $\theta_0$, as shown in Fig. 1.  For the system to jump to the
other state, the environment must be caught in a fluctuation which
points along the other direction, $-\theta_0$.  The amplitude for $N$
spin-1/2's to be found at an angle $2\theta_0$ away is
$\cos^N{\theta_0}$.  The new tunnel splitting is thus reduced by a
Franck-Condon type overlap factor: 
\be t^* =
t\<\theta_0|R_y(2\theta_0)|\theta_0\> = t \cos^N{\theta_0} = t
\frac{1}{(1+\lambda^2/\omega^2)^{N/2}} 
\ee 
where $R_y(2\theta_0)$ is the rotation operator. Note that this is
still a purely conservative process; the resonance is perfectly sharp
although at an exponentially reduced energy scale.  We will refer to
this resonance as the Franck-Condon resonance.

Although, for finite $N$ and $t$, the symmetry is unbroken and the
ground state is an evenly weighted superposition, there must be some
symptom of the crisis at $t \sim \lambda^2/2\omega$ when $N$ becomes
large.  Presumably, this symptom should be the inelastic broadening of
the resonance to an over damped state.  Following along the lines of
the Fermi Golden Rule (FGR) calculation in \cite{leggett_rmp}, we
treat the last two terms of (\ref{ham2}) as the unperturbed Hamiltonian,
$H_0$.  $H_0$ is a Zeeman-like \ha with a magnetic field that depends
upon the $z$-component of the system spin.  Therefore it should be
diagonalized with a $\sigma_z$ dependent rotation about the $y$-axis:
\be R \equiv R_y(\sigma_z\theta_0) = e^{-i {\sigma}_z \theta_0 {S}_y}
\ee The transformed \ha is then \be H_0^{\prime} = R^{\dagger}H_0 R =
\frac{\omega}{2N}S_z \cos{\theta_0} (1 + \frac{\lambda^2}{\omega^2})
\ee where $\tan{\theta_0}=\lambda/\omega$ has been chosen to eliminate
the $S_x$ term in $H_0^{\prime}$.  Now turning to the first term of
(\ref{ham2}), the rotation yields: \be H^{\prime}_1 \equiv \frac{t}{2}
R^{\dagger}\sigma_x R = \frac{t}{2}(\sigma^{+} e^{-i\sigma_z \theta_0
S_y} + \sigma^{-} e^{+i\sigma_z \theta_0 S_y}) \ee Now the system is
prepared in a state $|+\>$ with the environment in its ground state
$S_z=-N/2$ (which will be denoted $|-N/2\>$). The transition rate is
now calculated by applying Fermi's Golden Rule, treating
$H^{\prime}_1$ as the perturbation.  \be \Gamma =
\frac{t^2}{4}\sum_{m=-N/2}^{N/2}|\<-|\<m| \sigma^{-} 
e^{+i\sigma_z \theta_0 S_y}
|-N/2\>|+\>|^2 \delta(E_f-E_i) \ee

\begin{figure}[tb]
\epsfysize=3.0in
\centerline{\epsfbox{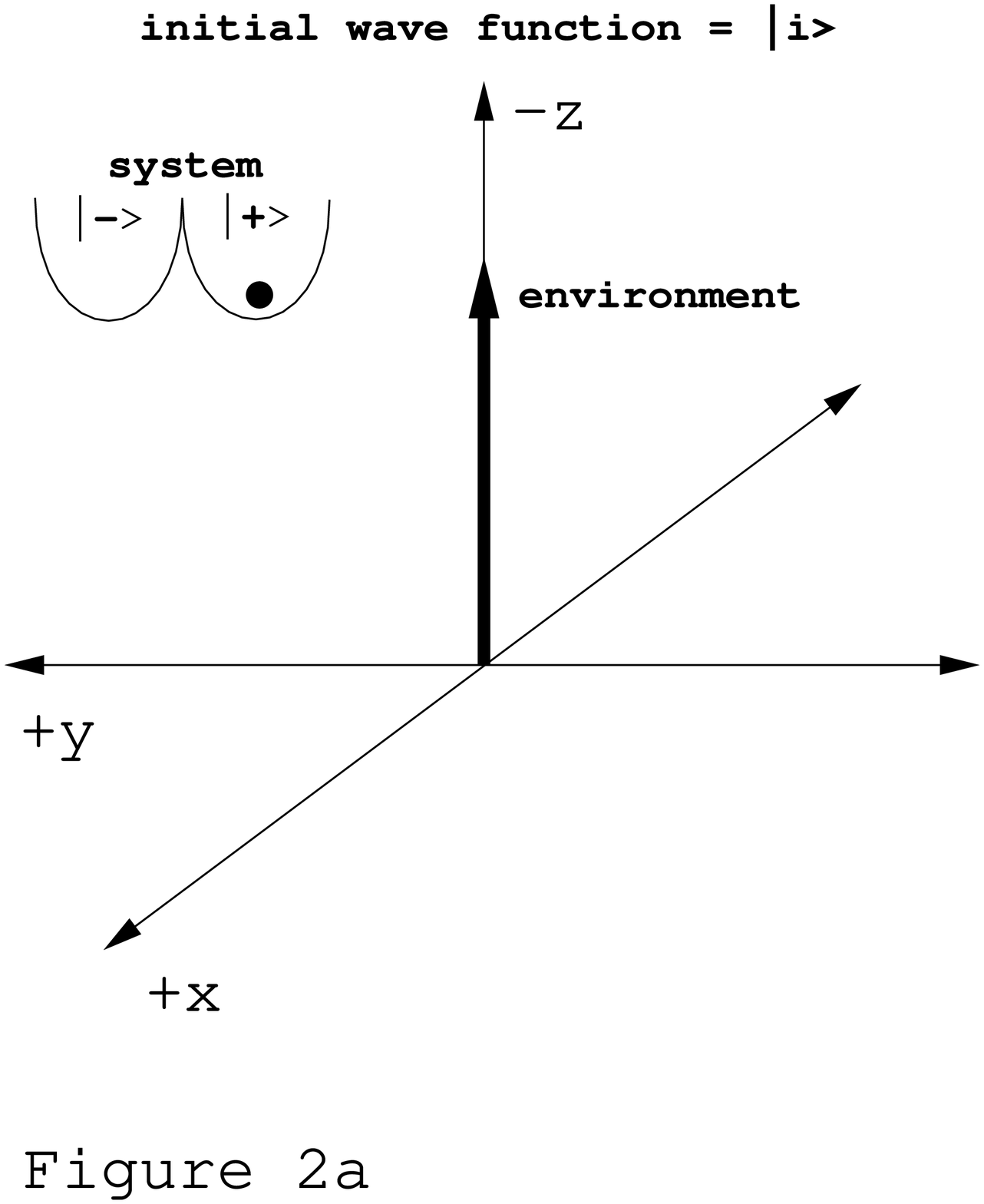}} 
\vskip 0.5truein
\epsfysize=3.0in
\centerline{\epsfbox{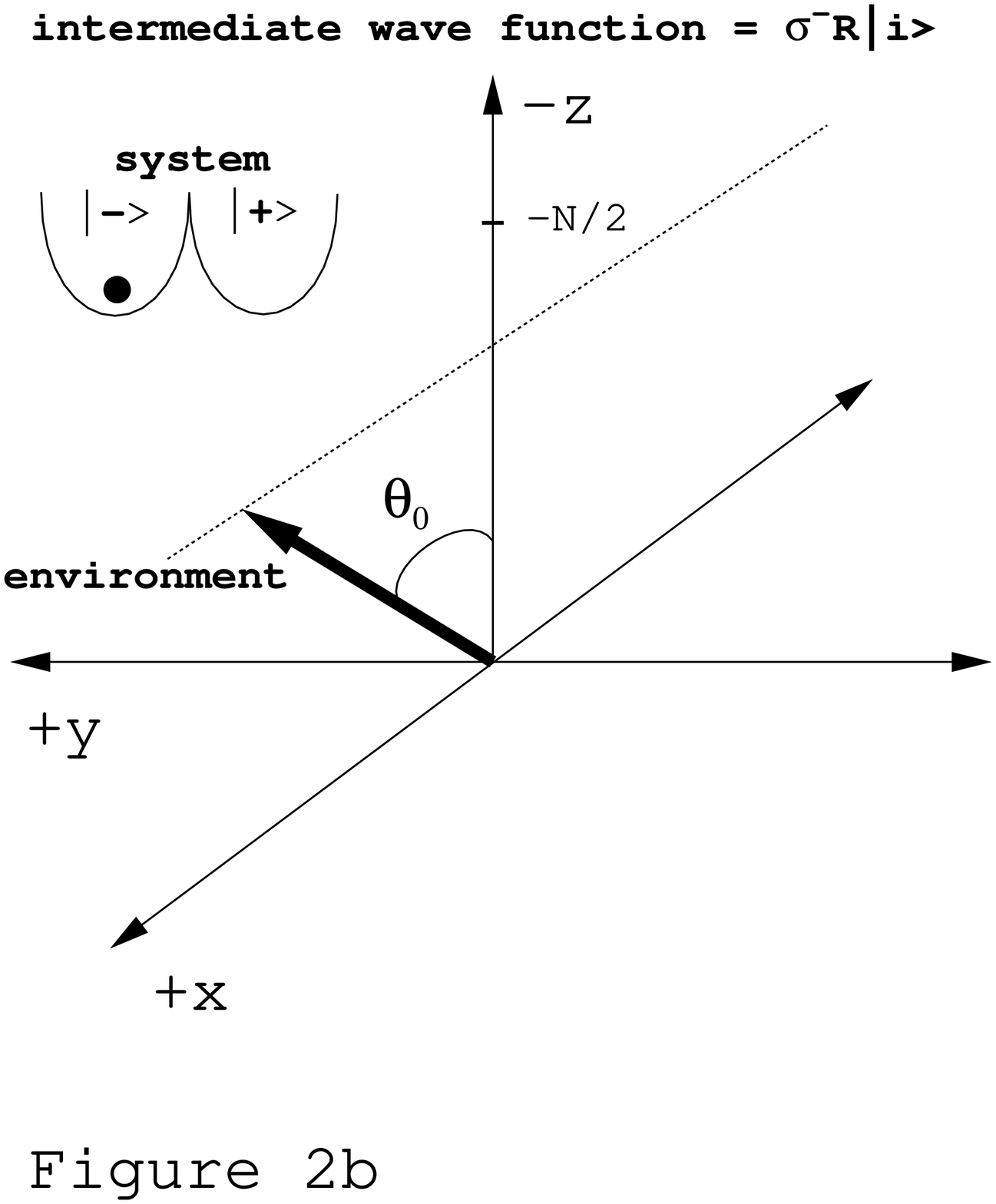}} 
\epsfysize=3.0in
\centerline{\epsfbox{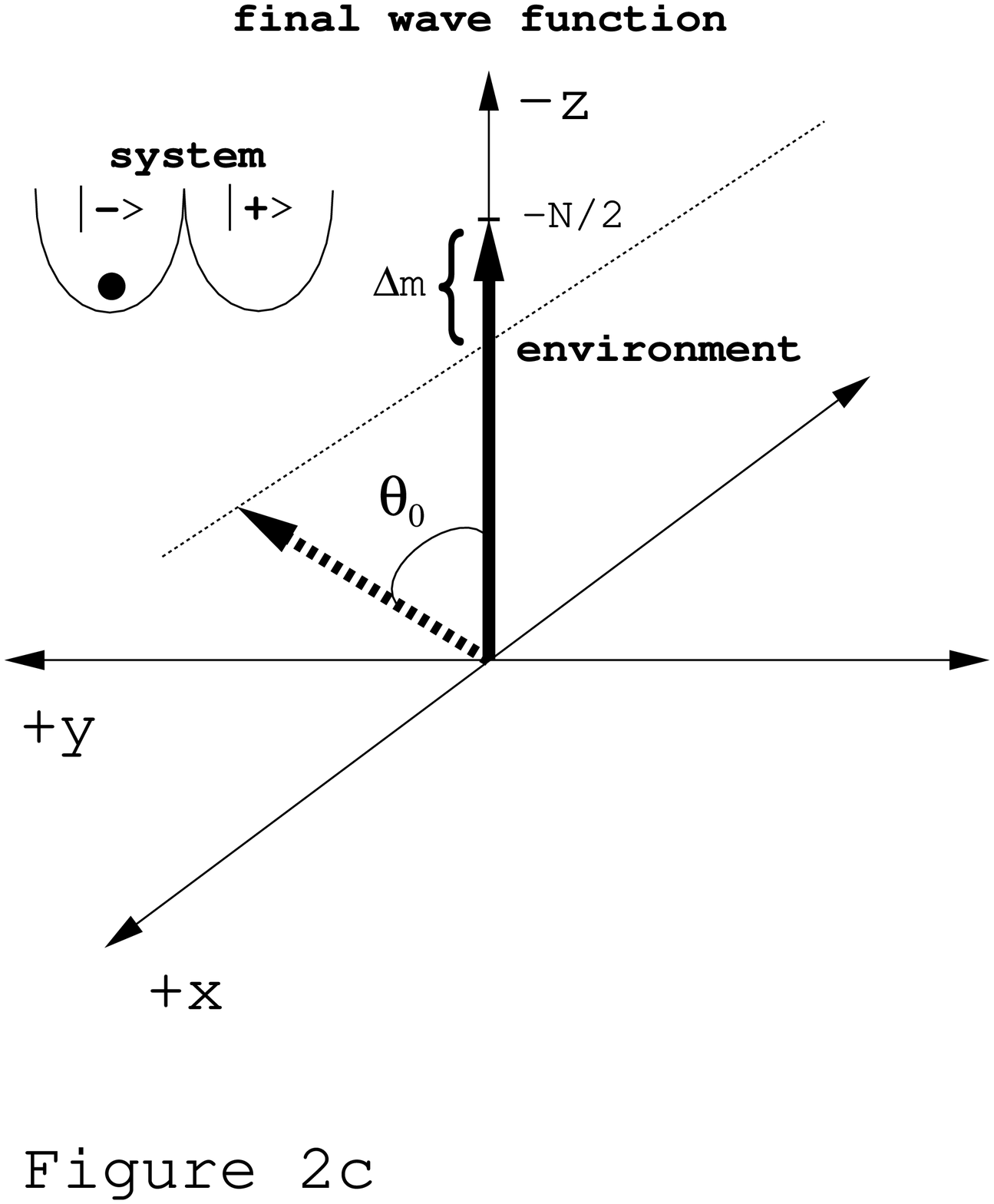}} 
\vskip 0.3truein \protect\caption{Sequence of states (a) initial, (b)
intermediate and (c) final, depicting Fermi's Golden Rule applied to
the Mermin Model. Solid black arrow is the environment; state of the
system is shown in the upper left. Final environment state must make a
fluctuation of approximately $\Delta m= -N/2\cos{\theta_0} + N/2$ to
have a reasonable overlap with the intermediate state. The system
drives the environment to dissipate energy.}
\end{figure}

As depicted in Figs. 2(a-c), $H^{\prime}_1$ induces a rotation of the
environment by $-\theta_0$ about the $y$-axis as the system flips from
$|+\>$ to $|-\>$. Since the environment is now rotated away from the
environment ground state of the {\em final} configuration
($\<-|\<-N/2|$) as well (Fig. 2c), the dominant contribution to $\Gamma$
comes from an excited (rotated) final state.  Looking at the geometry of
Fig. 2c, the final state, $\<-|\<m= -N/2 \cos{\theta_0}|$,
contributes most strongly.  Using the \ha $H_0^{\prime}$, the energy
difference is computed:
\be
\Delta E = E_f - E_i \sim \frac{\omega}{4}
(1-\cos{\theta_0})\cos{\theta_0}(1+\lambda^2/\omega^2)
\sim \frac{\lambda^2}{8\omega}
\ee
This calculation suggests an interpretation of the earlier result
(\ref{crit_point}) for the critical coupling obtained in the 
$N \rightarrow \infty$ limit. When the width of the resonance $\Delta
E$ becomes comparable to $t$, the resonance is over damped and quantum
coherence is destroyed.  

Although both the adiabatic computation and the FGR one are small $t$
perturbation theories and involve the same environment overlap
amplitude, their interpretations are quite different.  In the former,
the system in the $|+\>$ state leaves an ``imprint'' upon the
environment which, because it is nearly orthogonal to the imprint left
by system state $|-\>$, reduces the tunneling amplitude by an
exponentially small factor.  The resonance must therefore shift from
$O(t)$ to a smaller energy scale $O(te^{-N/2})$. In the latter case,
the system forces the environment to dissipate energy and the
resonance is broadened by $O(\lambda^2/\omega)$ but remains nominally
at $O(t)$.

\begin{figure}[tb]
\epsfxsize=3.5in
\centerline{\epsfbox{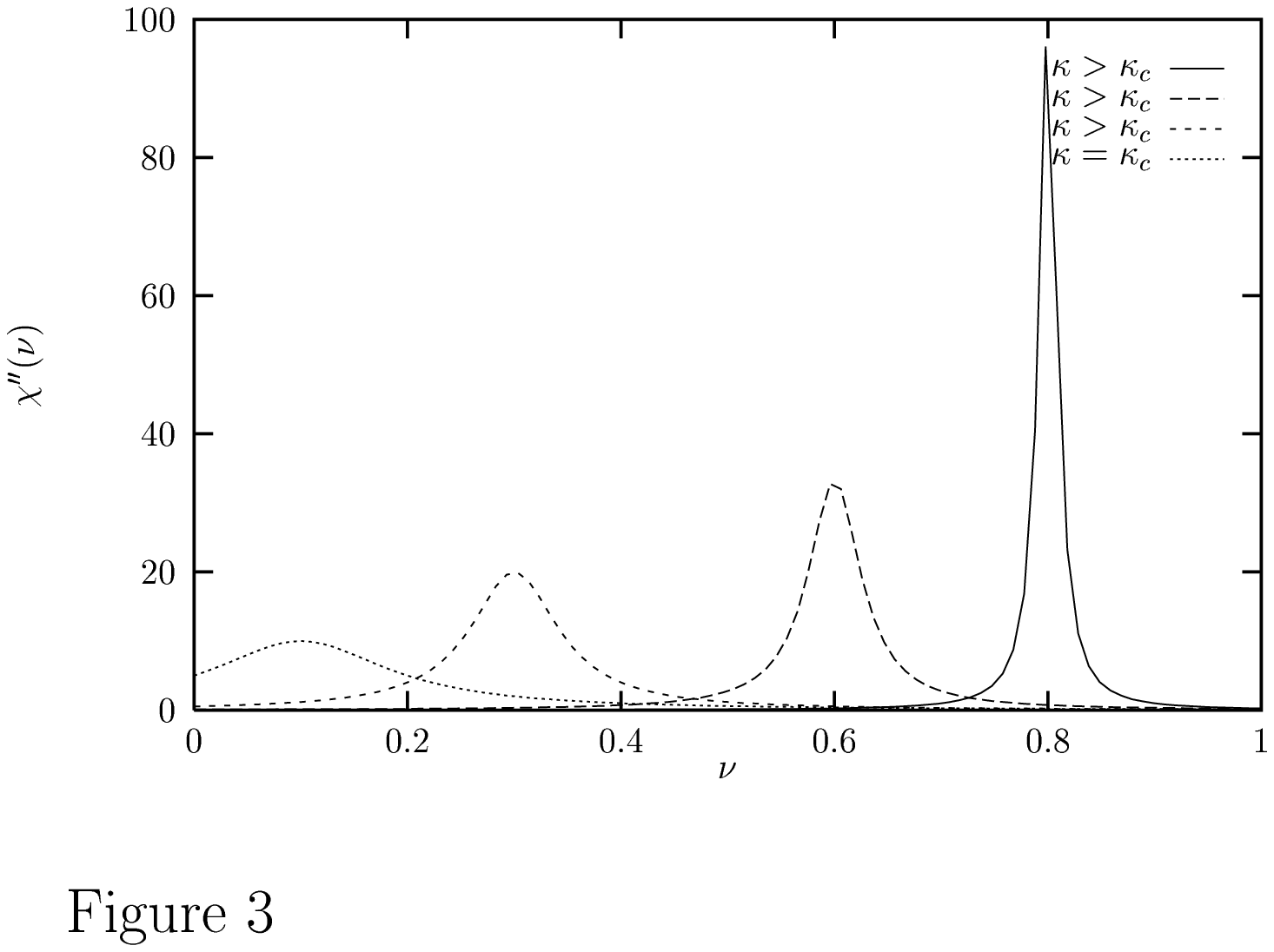}}
\vskip 0.5truein
\protect\caption{Possible ``continuous'' evolution of spectral weight.  The
principal resonance and the Franck-Condon resonance are the same.}
\end{figure}

One might speculate upon two possible scenarios in which these two
perturbative calculations are blended, as suggested in Figs. 3 and 4.
In Fig. 3, a single resonance broadens and shifts to lower energy,
decoherence occurring when the resonance becomes broadened beyond its
energy scale.  In this scenario, there is no explicit Franck-Condon
energy scale, just the appearance of increasing spectral weight close
to zero as the resonance becomes over-damped.  In the other scenario,
demonstrated in Fig. 4, spectral weight is only transferred to an
exponentially small energy below some critical coupling; the resonance
at $t$, although broadened, is not critically damped and remains
distinct from the Franck-Condon resonance.  To put it another way, the
systems retains a gap of $O(t)$.  Furthermore, even at finite $N$,
there is essentially no ``Franck-Condon effect'' for $\kappa >
\kappa_c$ (see Fig. 5). It is the second scenario that is
manifested in the Mermin Model.

\section{Calculations of the Linear Susceptibility: $\imchi$}
\label{lin_calculations}

We consider the \ha (\ref{ham}) with an additional coupling to an
external field $h(t) = h\cos{\nu t}$:
\be
\label{ham_ext}
H_{\mbox{ext}} = -h \frac{\sigma_z}{2} \cos{\nu t}
\ee
For calculating nonlinear \suss it will be necessary to use explicitly
real driving fields, so we adopt explicitly real notation for the
expectation value of the time dependent system spin, at this point.
The time dependence of $\sigma_z$, at linear response, is
\be 
\label{time_dep}
\half\<\sigma_z(t)\> = h \chi^{\prime}(\nu)\cos{\nu t} -
h \chi^{\prime\prime}(\nu)\sin{\nu t}
\ee
At zero temperature, the response function $\chi(\nu)$ is given by the
integral on the right hand side of (\ref{imchi}).  Expanded out,
\be
\label{chi1}
\chi(\nu) = \frac{1}{\hbar}\sum_{n}{\frac{1}{4} |\sigma_{0n}|^2
(\frac{1}{\nu + \nu_{n0} - i\delta} - 
\frac{1}{\nu - \nu_{n0} - i\delta})}
\ee
where $\nu_{ij} \equiv (E_i - E_j)/\hbar$ and $\delta = 0^+$.  The
matrix elements of the system spin $\sigma_{ij} \equiv
\<i|\sigma_z|j\>$ will be referred to as the order parameter
matrix. (The state $|i\>$ now refers to the combined system and
environment state.)  Using (\ref{time_dep}), the rate of energy
absorption is then
\be
\label{P1}
\Omega = -\half \nu \imchi h^2
\ee
Equations analogous to (\ref{time_dep}), (\ref{chi1}) and (\ref{P1}) for the
third order nonlinear \sus are derived in the Appendix.

The \ha (\ref{ham}) was diagonalized and the complete order parameter
matrix, $\sigma_{ij}$, necessary for the linear and nonlinear
calculations, was computed.  It is natural to work with one
dimensionless coupling constant, $\kappa \equiv 2\omega t/\lambda^2$.
The critical coupling implied in the $N \rightarrow \infty$
calculation is then $\kappa = \kappa_c = 1$.

\begin{figure}[tb]
\epsfxsize=3.5in
\centerline{\epsfbox{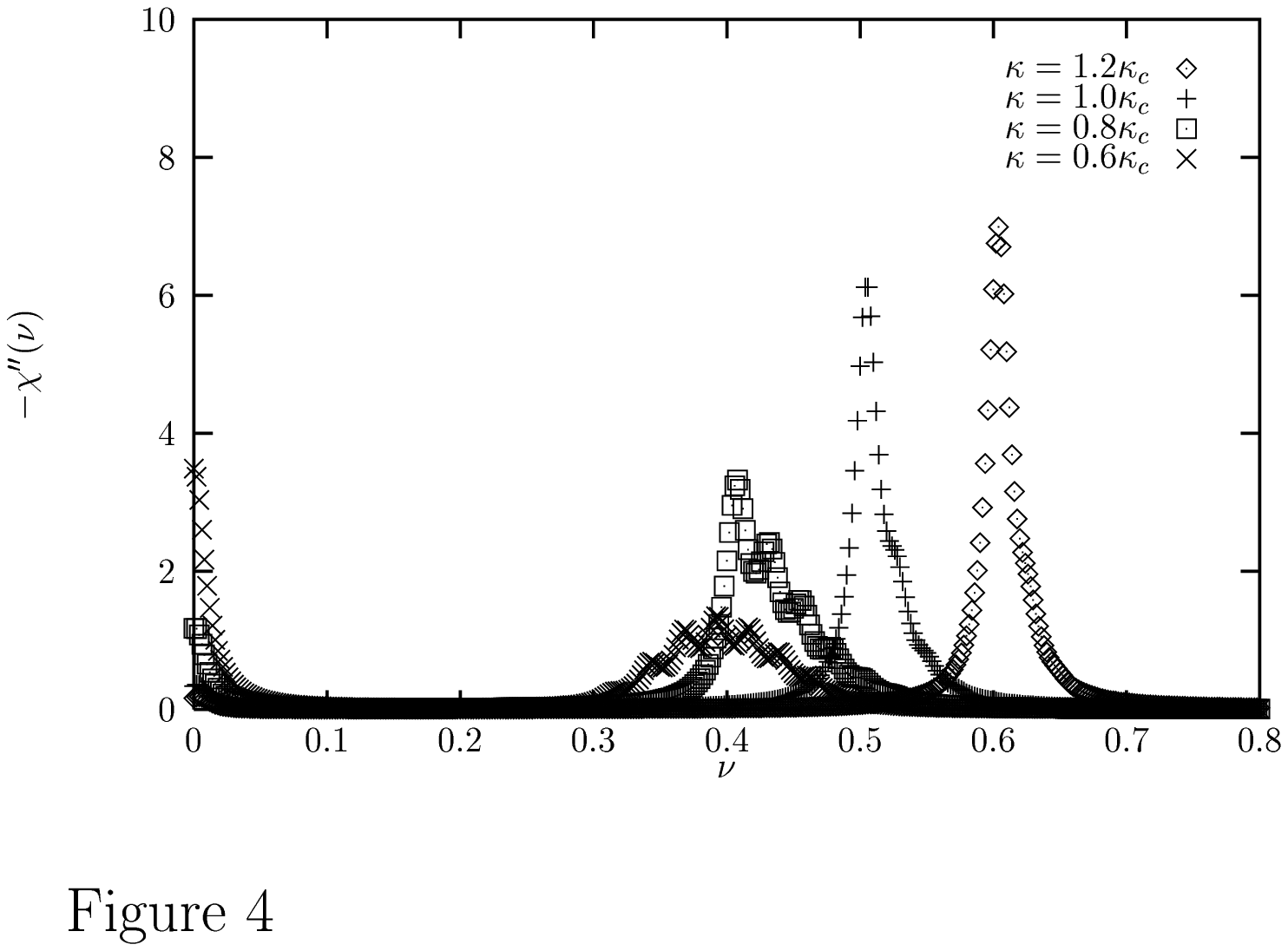}}
\vskip 0.5truein
\protect\caption{Evolution of spectral weight in the Mermin Model.  $\imchi$
is shown for four values of $\kappa$ decreasing below the critical
value, $\kappa_c$.  Despite some change in the position of the
principal resonance, spectral weight emerges at $\nu \sim 0$ abruptly
at $\kappa_c$ and the two features remain distinct.}
\label{linear_sus}
\end{figure}

Fig. 4 shows the behavior of $\imchi$ as the system energy scale $t$
is reduced, corresponding to a range $\kappa = 0.6\kappa_c -
1.2\kappa_c$.  These computations were performed for $80$ environment
spins.  Computations with up to $300$ environment spins suggest that
there is little change beyond $80$ spins.  An artificial broadening
($\delta=0.01$) was added to make the features more visible. As seen
in this sequence, $\kappa=\kappa_c$ is marked by the appearance of the
Franck-Condon resonance at an exponentially small energy
scale ($\nu_{10} \sim 10^{-3}$).  The spectral weight remaining at
$O(t)$ when $\kappa > \kappa_c$ clearly exhibits inelastic broadening,
although the resonance essentially disappears before becoming
critically damped.

\begin{figure}[tb]
\epsfxsize=3.5in
\centerline{\epsfbox{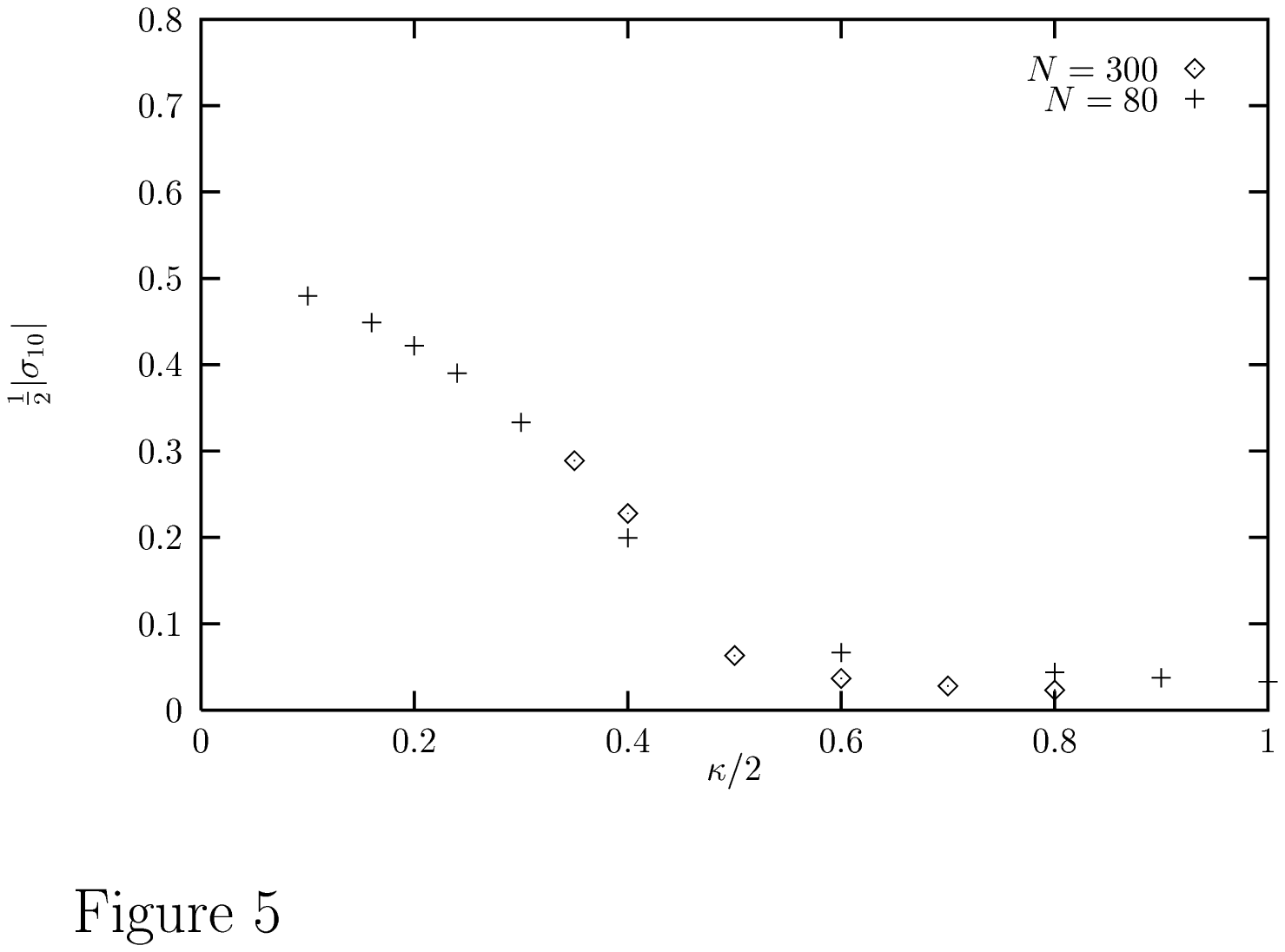}}
\vskip 0.5truein
\protect\caption{The order parameter, $\sigma_{10}$, is plotted as a function
of $\kappa$ and exhibits second order phase transition behavior.
$\kappa$ was varied by changing $t$.  Comparison of $N=80$ and $N=300$
data show the slope steepening at the critical point.  When $\kappa >
\kappa_c$, there is no weight in the Franck-Condon resonance other
than that attributed to the rounding of the phase transition at finite
$N$.}
\end{figure}

The counterpart, at finite $N$, to $\<\sigma\> \neq 0$ at infinite $N$
is the spectral weight at the Franck-Condon resonance,
$\sigma_{10} \neq 0$.  Fig. 5 demonstrates that $\sigma_{10}$ behaves
as expected in a second order phase transition; the slope at $\kappa =
\kappa_c^+$ grows with increasing $N$.  The matrix, $\sigma_{ij}$, and
consequently, $\imchi$, obey a sum rule: 
\be 
\label{sum_rule}
\sum_i{|\sigma_{ij}|^2} = 1 \ee An incompletely developed broken
symmetry ($|\sigma_{10}|^2 < 1$) means that spectral weight remains at
the principal resonance at $O(t)$.  This behavior is somewhat analogous to the
superconducting transition and the Glover-Tinkham-Ferrell sum rule for the
real part of the dynamical conductivity, $\sigma(\omega)$: \be \int_0^\infty
\sigma_{\mbox n}(\omega) d\omega = \frac{\pi n_{\mbox s}e^2}{2m} +
\int_\Delta^\infty \sigma_{\mbox s}(\omega) d\omega \ee (Subscripts
``n'' and ``s'' refer to ``normal'' and ``superconducting''.) The sum
rule states that spectral weight missing from dynamical conductivity
above the gap must reappear as superfluid (DC) Drude weight.  This
behavior may be compared with other models and computations of
environmental spin decoherence \cite{garg_stamp_levine}.  These models
differ from the Mermin Model in two respects: 1) typically the
environmental coupling constant is comparable to the system energy
scale instead of weak $O(\lambda/N)$ coupling, as in the present case;
2) the coupling between the system and environmental spins is
isotropic, of the form, $\vec \sigma \cdot \vec s_j$.  These
differences lead to qualitatively different decoherence effects.

\section{Nonlinear Dynamical Susceptibility}
\label{nonlin_calculations}

In studying the nonlinear susceptibilities, we are interested in power
absorption, as this is the relevant experimental probe for MQC.
Consider the response of just a single two-level-system (TLS) coupled to
ancillary levels: Driving the system with a signal at frequency $\nu$
results in third order nonlinear responses at $\nu$ and $3\nu$.  The
signal at $3\nu$ is not absorptive and, in any case, would not be
detected in any experimental setting where pumping and detection take
place at (nearly) the same frequency.  If the nominal frequency scale
of the TLS is $\nu_0$, there is an imaginary but {\em emissive}
response at $\nu_0$.  (The non vanishing, odd order, nonlinear
corrections to power absorption at $\nu_0$ must alternate in sign to
produce finite absorption at $\nu_0$, consistent with the exact Rabi
solution.)  On the other hand, third order two-photon absorption (TPA)
process at $\nu_0/2$ would resemble ordinary linear absorption of a
$\nu_0/2$ TLS in that it would be detected as an absorptive phase
shift when the pumping frequency reaches $\nu_0/2$.  The reader is
referred to ref. \cite{butcher} for further details about nonlinear
processes.

Fig. 6 shows the two photon absorption spectrum along with the linear
susceptibility for several values of $\kappa$.  Our basic result is
the following: as the critical value of $\kappa$ is reached and the
two-level-system undergoes a transition to decoherence, the two photon
absorption power, $\Omega_{2\gamma}$, begins to increase.  It reaches a
maximum at approximately $\kappa=\kappa_c/2$ and then decreases to
zero as $\kappa$ goes to zero.  Fig. 7 shows the behavior of the TPA
signal as a function of the control parameter $\kappa$.  

The two photon absorption spectrum was calculated from an exact
diagonalization of the Mermin model using the following expression for
TPA (which is derived in the Appendix):
\begin{eqnarray}
\label{tpa}
\Omega_{2\gamma} = && \frac{1}{2}h^4 \nu \chi^{[3]}(\nu) \nonumber
\\ = && \frac{1}{8} (\frac{h}{2})^4 \nu \frac{1}{\hbar^3}
\sum_{nml}{\sigma_{0n}\sigma_{nm}\sigma_{ml}\sigma_{l0}} \nonumber \\
&& \times 
P_{0m}^{\prime \prime}(-2\nu)
P_{l0}^{\prime}(\nu)
P_{n0}^{\prime}(\nu)
\end{eqnarray}
In the expression above, the energy denominators have been abbreviated,
\be
P_{ij}(\nu) \equiv \frac{1}{\nu - \nu_{ij} -i\delta}
\ee
To calculate (\ref{tpa}) it is necessary to find all possible
sets of four couplings that connect the ground state back to the
ground state.  These are referred to as ``four link chains''; a few
such chains are depicted in Fig. 8. 

\begin{figure}[tb]
\epsfxsize=3.5in
\centerline{\epsfbox{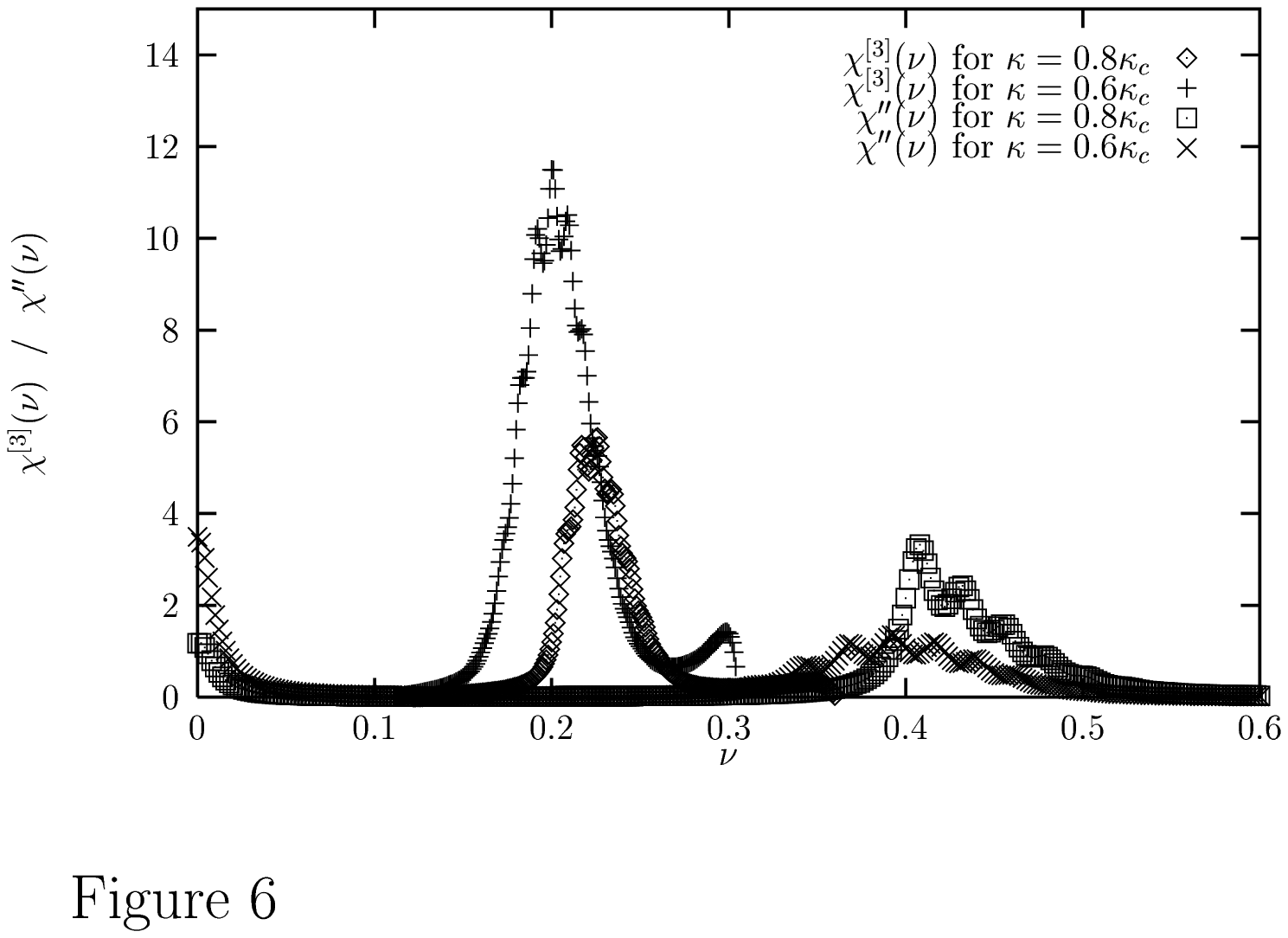}}
\vskip 0.5truein
\protect\caption{Nonlinear \suss for two photon absorption compared with the
linear \sus for two values of $\kappa$.  Note the increase in
$\chi^{[3]}$ as $\kappa$ is decreased from $0.8\kappa_c$ to
$0.6\kappa_c$.}
\end{figure}

To understand the non-monotonic behavior of the TPA signal, we look at
the expression (\ref{tpa}).  Isolating the absorptive pole,
$\chi^{[3]}(\nu)$ can be factored in the following way: 
\be
\chi^{[3]}(\nu) = \frac{1}{4} (\frac{1}{2})^4 \frac{1}{\hbar^3}
\sum_{m}{P_{0m}^{\prime \prime}(-2\nu)|\alpha_m(\nu)|^2} 
\ee 
where $\alpha_m(\nu)$ is a sum over intermediate states, 
\be
\label{alpha}
\alpha_m(\nu)
\equiv \sum_n{\sigma_{0n}\sigma_{nm}P_{n0}^{\prime}(\nu)} =
\sum_n{\frac{\sigma_{0n}\sigma_{nm}}{\nu - \nu_{n0}}} 
\ee
$\alpha_m(\nu)$ is known in nonlinear optics literature as the
degenerate hyperpolarizability. We focus our attention upon the {\sl
integrated} spectral weight, shown below for both linear and nonlinear
susceptibilities $(\hbar = 1)$:
\begin{equation}
I_{\mbox{lin}} = \int_{0^+}^{\infty}{\chi^{\prime \prime}(\nu) d\nu}
~~~~~~~
I_{\mbox{nonlin}} = h^2\int_{0^+}^{\infty}{\chi^{[3]}(\nu) d\nu}
\end{equation}
The limits of integration are meant to restrict the integration to the
principal resonance and exclude the contributions from the near
degenerate ground states.  (These are the states $m=0,1$ excluded in
the sum below.)  The integrated weight is in some sense a
``figure of merit'' for detection of an MQC resonance.  A meaningful
comparison between linear and nonlinear absorption requires the
multiplicative factor $h^2$ in the nonlinear expression.  In both
cases, the integration collapses the absorptive pole leaving, in the
linear case,
\be
\label{isw_lin}
I_{\mbox{lin}} = \pi (\half)^2 \sum_{n \neq 1}{|\sigma_{0n}|^2} =
\pi (\half)^2(1-\sigma_{01}^2)
\ee
and in the nonlinear case,
\be
\label{isw_nonlin}
I_{\mbox{nonlin}} = \pi h^2(\half)^4 \frac{1}{8} \sum_{m \neq 0,1}
 {|\alpha_m(\nu_{m0})|^2}
\ee

\begin{figure}[tb]
\epsfxsize=3.5in
\centerline{\epsfbox{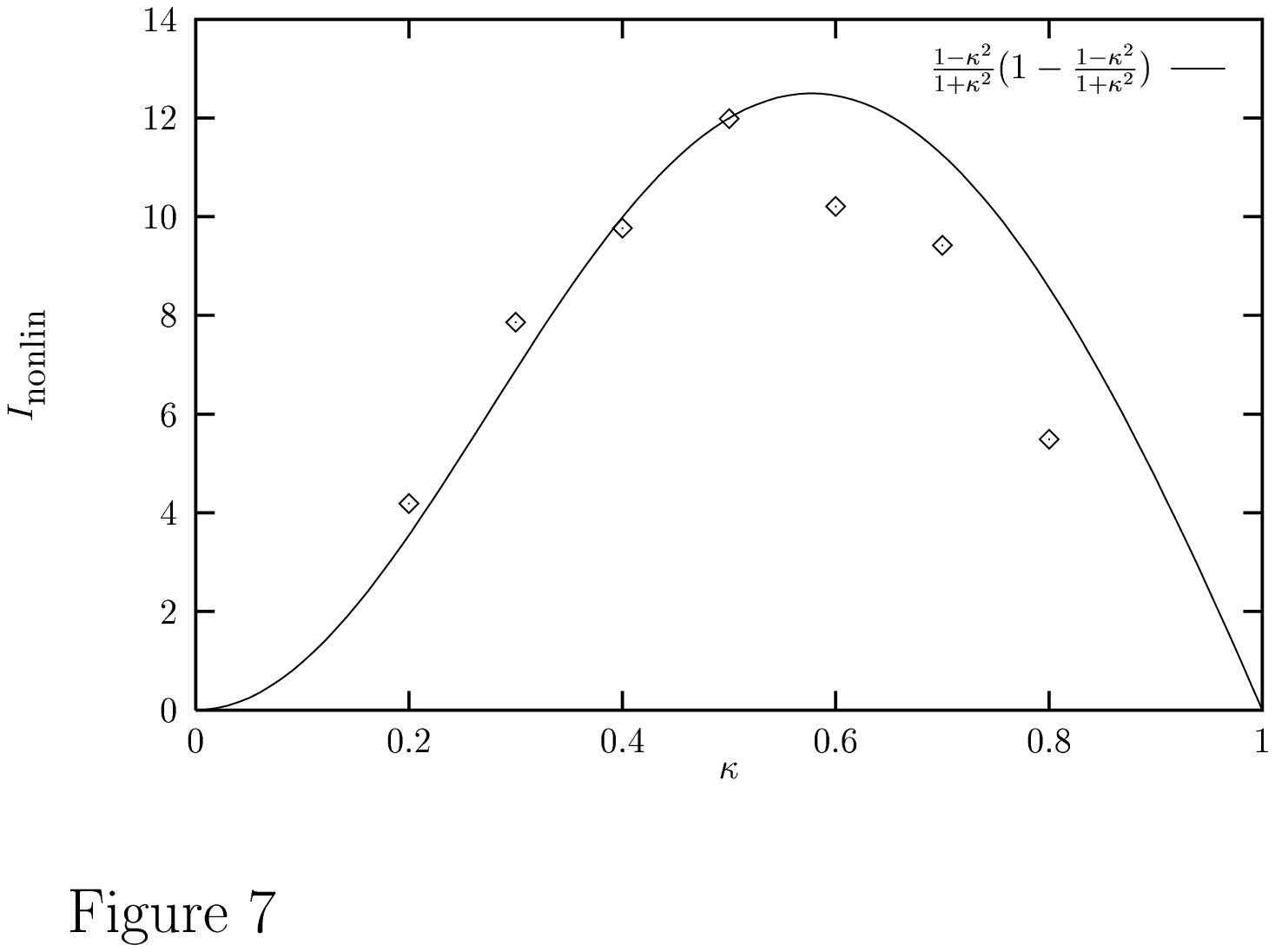}}
\vskip 0.5truein
\protect\caption{Integrated spectral weight $I_{\mbox{nonlin}}$ for two photon
absorption as a function of $\kappa$.  Solid line is the result of
theory.  The applied AC field strength $h=1$.}
\end{figure}

Since $I_{\mbox{lin}}$ given in equation (\ref{isw_lin}) depends only
upon the order parameter $\sigma_{01}$, it has a universal behavior in
terms of $\kappa$ through the decoherence transition (see equations
(\ref{sigma_exact_sol}), (\ref{i_lin})). It turns out to be possible to 
express $I_{\mbox{nonlin}}$ in a similar fashion within a simple
approximation and understand the non-monotonic behavior seen in
fig. 7.  It will be shown that $I_{\mbox{nonlin}}$ depends upon the
order parameter through:
\be
\label{isw_nonlin2}
I_{\mbox{nonlin}} = \frac{2 \pi h^2}{\nu_0^2}(\half)^4 \sigma_{01}^2
(1 - \sigma_{01}^2) + O(\frac{h^2\epsilon}{\nu_0^3})
\ee
and thus reaches a maximum at approximately $\kappa = \kappa_c/2$.
(In the exact solution, $|\<0|\sigma_z|0\>|^2 \sim 1-\kappa^2$ for
$\kappa_c=1$; see equation (\ref{sigma_exact_sol}).) 

To understand this result, consider the usual conditions through which
a broken symmetry is accessed mathematically: Apply a small static
bias field $h_{{\rm dc}}$ coupled to the system through
$H_{{\rm dc}} \equiv -h_{{\rm dc}} \sigma_z$ and take the limits in
the order,
\be
\lim_{h_{\rm dc} \to 0} \lim_{N \to \infty} \<0|\sigma_z|0\>
\ee
In the decoherent phase ($\kappa < \kappa_c$), a field $h_{{\rm dc}}
\sim O(te^{-N/2})$ will localize the system yielding $\<0|\sigma|0\>
\ne 0$, but will have essentially no effect in the coherent phase 
($\kappa > \kappa_c$). The relevant energy levels now look like the
right half of Fig. 8 (the quasi-degenerate ground state levels are no
longer coupled through $\sigma_z$).  

\begin{figure}[tb] 
\epsfxsize=3.5in 
%
\centerline{\epsfbox{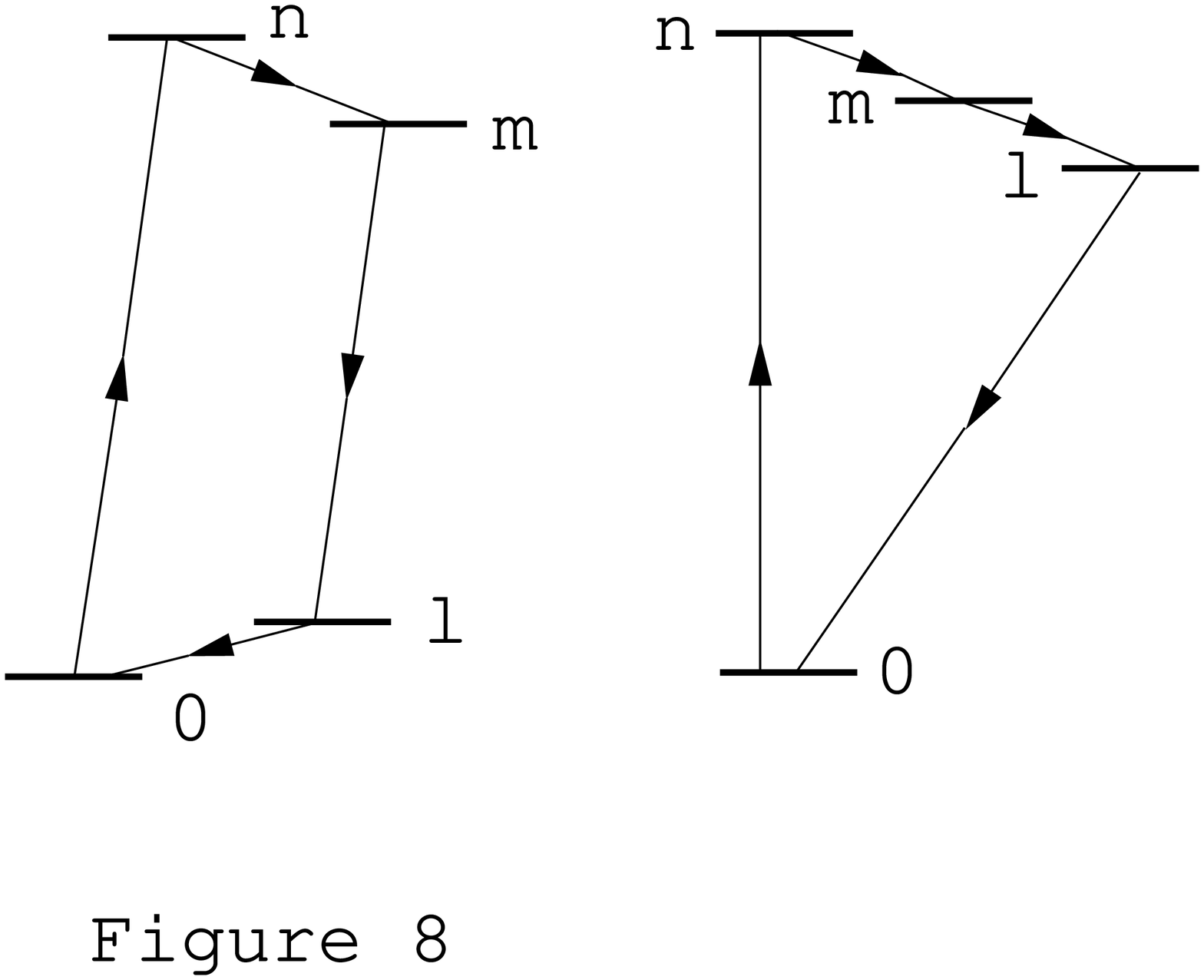}} 
\vskip 0.3truein \protect\caption{Two possible four-link-chains.}
\end{figure} 

The key observation is that the net transition amplitude (proportional
to $\alpha_m(\nu)$) to go from $|0\>$ to a state $|m\>$ through
degenerate intermediate states $|n\>$, $|l\>$, etc. is zero {\sl
unless there is a broken symmetry} leaving $\<0|\sigma_z|0\> \ne 0$.  If
the intermediate levels are degenerate, $\alpha_m(\nu)$ may be written
\begin{displaymath}
\alpha_m(\nu) =
\sum_{n \ne 0,m}{\frac{\sigma_{0n}\sigma_{nm}}{\nu - \nu_{n0}}} =
\frac{1}{\nu - \nu_{n0}}\sum_{n \ne 0,m}{\sigma_{0n}\sigma_{nm}}= \nonumber
\end{displaymath}
\be
= -\frac{1}{\nu - \nu_{n0}}
(\<0|\sigma_z|0\>\<0|\sigma_z|m\> + \<m|\sigma_z|m\>\<0|\sigma_z|m\>)
\ee
where the last line is obtained by inserting a complete set of states
into $\<0|\sigma_z^2|m\> = 0$.  Since the expectation value of $\sigma_z$
is $O(1)$ only in the ground state (in the broken symmetry phase), we
arrive at
\be
\alpha_m(\nu) = -\frac{\<0|\sigma_z|0\>\<0|\sigma_z|m\>}{\nu - \nu_{n0}}
= -\frac{\sigma_{00}\sigma_{0m}}{\nu - \nu_{n0}}
\ee
Unless $\sigma_{00}$ is non zero, the transition amplitudes for $|0\>
\rightarrow |m\>$ involving an intermediate state interfere
destructively.  Finally, the integrated weight is obtained from
(\ref{isw_nonlin}),
\begin{eqnarray}
I_{\mbox{nonlin}} = && {\rm const} \times \sum_{m \neq 0}
 {|\alpha_m(\nu_{m0})|^2} \nonumber \\  
= && \frac{{\rm const}}{(\nu - \nu_{n0})^2}
\times \sum_{m \neq 0}{\sigma_{00}^2\sigma_{0m}\sigma_{m0}} \nonumber \\
= && \frac{{\rm const}}{(\nu - \nu_{n0})^2} \times 
\sigma_{00}^2 (1 - \sigma_{00}^2)
\end{eqnarray}
where the sum rule (\ref{sum_rule}) (with $j=0$) was used in the last
line.  This is the argument used to establish the result
(\ref{isw_nonlin2}).

We now fill in the details of this argument appropriate for the
numerical computation performed in this work by 1) setting $h_{\rm
dc}=0$ and 2) including the finite width of the resonance.
$\alpha_m(\nu)$ is to be evaluated at the frequency for TPA, $\nu =
\nu_{m0}/2$---this frequency is about half of the typical frequencies
of the intermediate levels $\nu_{n0}$. $\alpha$ is rewritten in such a way
as to separate the quasi-degenerate ground states from the excited
states at the energy of the principal resonance at $O(t)$ (see Fig. 9):
\be
\label{alpha2}
\alpha_m(\nu_{m0}/2) = \frac{\sigma_{01} \sigma_{1m}}{\nu_{m0}/2 -
\nu_{10}} + \sum_{n \neq 0,1}{\frac{\sigma_{0n}
\sigma_{nm}}{\nu_{m0}/2 - \nu_{n0}}} 
\ee 
$\nu_{10} = t^* \sim O(te^{-N/2})$ is the energy scale of the
Franck-Condon resonance and will be set to zero.  The
difference between the resonant frequency, $\nu_{m0}$, and the
intermediate state level, $\nu_{n0}$, which appears in the second term
of (\ref{alpha2}) is bounded by the width of the primary resonance and
is therefore typically much smaller than the primary resonance
itself.  Denoting a maximum value for that difference by $\epsilon$
(that is, $\nu_{n0} = \nu_{m0} + \epsilon$), the second term of
(\ref{alpha2}) becomes:
\be
\sum_{n \neq 0,1}{\frac{-\sigma_{0n}
\sigma_{nm}}{\nu_{m0}/2}} + O(\frac{\epsilon}{\nu_{0}^2})
\ee
where $\nu_{0}$ is the nominal energy scale of the resonance.
Following the previous calculation for $h_{\rm dc} \ne 0$, the
hyperpolarizibility $\alpha$ is now
\be \alpha_m(\nu_{0}) =
\frac{4\sigma_{01}\sigma_{1m}}{\nu_{0}} +
O(\frac{\epsilon}{\nu_{0}^2}) 
\ee 
Substituting $\alpha$ back into equation (\ref{isw_nonlin}) and using
the sum rule (\ref{sum_rule}) (with $j=1$) establishes the result
(\ref{isw_nonlin2}).

To express $I_{\mbox{lin}}$ and $I_{\mbox{nonlin}}$ in terms of the
control parameter $\kappa$, we appeal to the exact solution in the $N
\rightarrow \infty$ limit.  Specifically, the matrix element
$\sigma_{01}$ is related to the expectation value of $\sigma_z$ in the
either of the degenerate ground states of (\ref{ham3}) (for $\kappa <
\kappa_c$) by:
\be
\label{sigma_exact_sol}
\lim_{N \rightarrow \infty} |\sigma_{01}| = |\<\sigma_z\>_0| = 
\sqrt{\frac{1-\kappa^2}{1+{\frac{\lambda^2}{\omega^2} \kappa^2}}} = 
\sqrt{\frac{1-\kappa^2}{1+{\frac{2t}{\omega} \kappa^2}}}
\ee
Therefore $I_{\mbox{lin}}$ and $I_{\mbox{nonlin}}$ may be compactly
written in terms of $\kappa$:
\be
\label{i_lin}
I_{\mbox{lin}} = \pi (\half)^2(1-\frac{1-\kappa^2}
{1+{\frac{\lambda^2}{\omega^2} \kappa^2}})
\ee
\be
\label{i_nonlin}
I_{\mbox{nonlin}} = \frac{2 \pi h^2}{\nu_0^2}(\half)^4
\frac{1-\kappa^2}
{1+{\frac{\lambda^2}{\omega^2} \kappa^2}}
(1-\frac{1-\kappa^2}
{1+{\frac{\lambda^2}{\omega^2} \kappa^2}})
\ee

\begin{figure}[tb] 
\epsfxsize=3.5in 
%
\centerline{\epsfbox{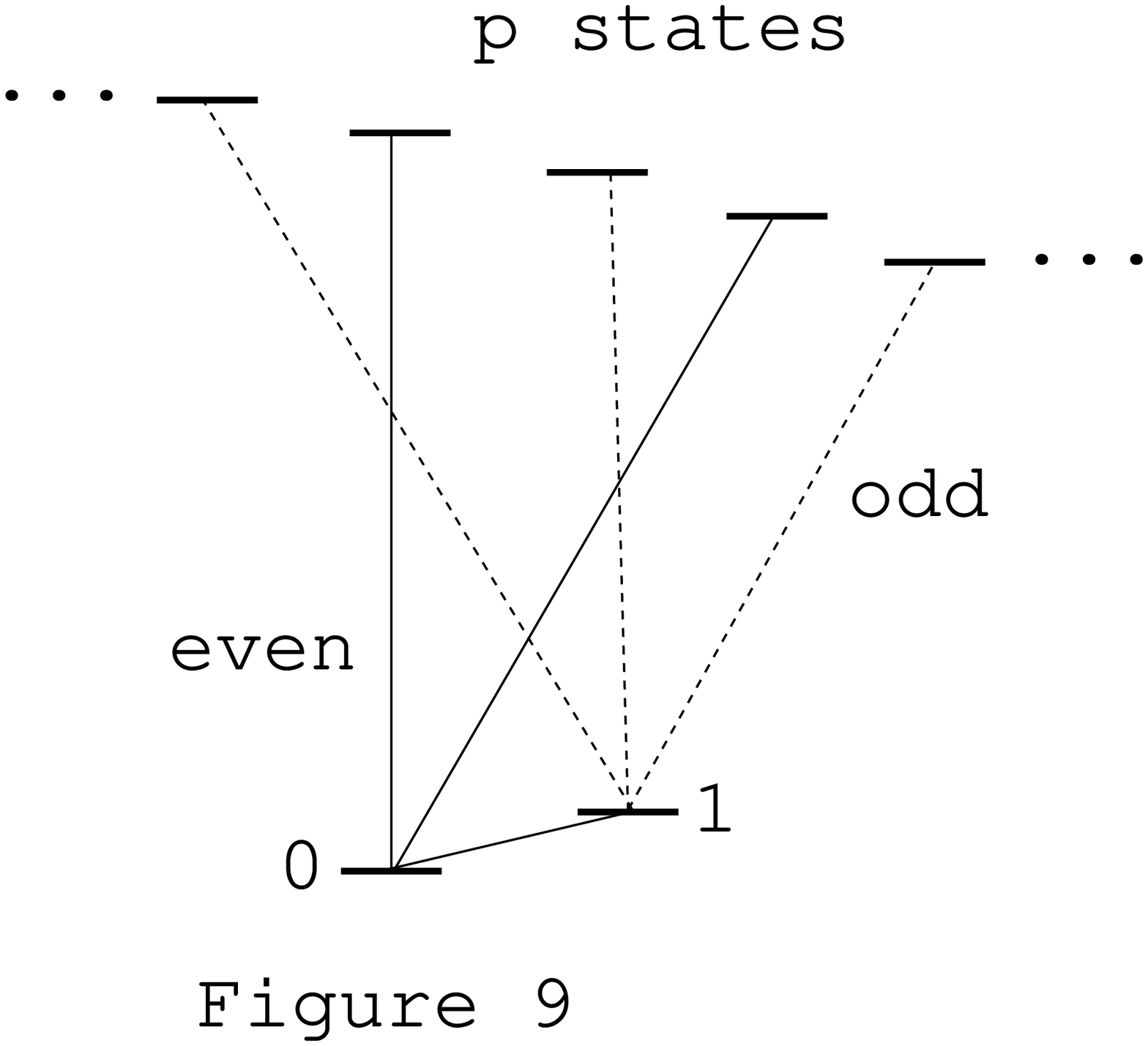}}
\vskip 0.3truein \protect\caption{Depiction of the quasi-degenerate
ground state $|0\>, |1\>$ coupling to $p$ excited states.  The
splitting between $|0\>, |1\>$ defining the Franck-Condon resonance is
$O(te^{-N/2})$.  The set of closely spaced excited states represents
the inelastic broadening of the principal resonance.}
\end{figure} 

Returning to fig. 7, the non-monotonic behavior of $I_{\mbox{nonlin}}$
as a function of $\kappa$ follows from its dependence upon the
combination: $\<\sigma_z\>^2(1-\<\sigma_z\>^2)$.  A plot of
$I_{\mbox{nonlin}}$ following from the analytic expression
(\ref{i_nonlin}) is shown for comparison.  In the numerical
computations, the maximum intensity appears at approximately $\kappa =
1/2$ in agreement with theory.

\section{Conclusion}
\label{conclusion}

In this work we have attempted to make more explicit the relationship
between the decoherence transition in the Mermin Model and a typical
thermodynamic phase transition. At the onset of decoherence, spectral
weight is transferred from above the ``gap,'' set by the system energy
scale $t$, to a delta function at $\nu=0$.  This behavior is in
contrast to the SBH where symmetry breaking is prefaced, first by the
complete disappearance of the inelastic peak in $\imchi$ at finite
frequency (when $\alpha=1/2$), and then finally by the softening of
renormalized system energy scale to zero (when $\alpha=1$).

Also, we have studied nonlinear absorption in a decoherent TLS.
Single photon absorption is a measure of coherence of the system
because it samples the spectral weight remaining at the principal
resonance---or, the complement of the order parameter according to the
sum rule.  On the other hand, two-photon-absorption (TPA) samples the
product of the order parameter and its complement and thus vanishes in
the fully coherent or fully decoherent phase.  TPA reaches maximum
intensity somewhere between these two extremes.

Observing TPA may be possible in systems demonstrating quantum
coherence such as the recently studied Fe$_8$ molecular nanomagnet.
\cite{delbarco98}.  In principal, measurement of the ratio of
$I_{\mbox{nonlin}}$ to $I_{\mbox{lin}}$ would provide a direct measure
of the order parameter $\<0|\sigma_z|0\>$, the localized ``weight'' of
the system.  If we consider a sample in which the system spins are
individually partially localized but uncorrelated with each other
(like a spin glass), TPA may be a useful probe for this type of order.

\acknowledgments

The author gratefully acknowledges the support of the Cottrell
Foundation through Research Corporation Grant number CC3834.  Work
performed at BNL supported by the U.S. DOE under contract no. DE-AC02
98CH10886.  The author also wishes to acknowledge many useful
discussions with Dr. Jeffrey Clayhold, Dr. Barry Friedman,
Dr. V. N. Muthukumar and the help of Joshua Eisner in performing some
preliminary computations.

\newpage

\appendix
\section{Derivation of the Nonlinear Dynamical Susceptibility}
\label{appendix}

We begin by expressing the full time dependence of the expectation
value of the desired operator in the presence of the driving
field.
\be
\label{S_matrix}
\<\sigma_z(t)\> = \<0|U^{\dagger}(t,-\infty) \sigma_z(t)
U(t,-\infty)|0\>
\ee
where

\begin{eqnarray}
U(t,-\infty) = &&{\mbox{\bf T}} e^{-i\int_{-\infty}^t
{H(t^{\prime})dt^{\prime}}}\nonumber \\
= &&1\underbrace{-i\int_{-\infty}^t{dt_1 H(t_1)}}_{I_1} \nonumber \\
&&\underbrace{- \int_{-\infty}^t{dt_1} \int_{-\infty}^{t_1}
{dt_2  H(t_1) H(t_2)}}_{I_2}\nonumber \\
&&\underbrace{+i\int_{-\infty}^t{dt_1} \int_{-\infty}^{t_1}{dt_2} 
\int_{-\infty}^{t_2}{dt_3  H(t_1) H(t_2) H(t_3)}}_{I_3} \nonumber \\
&& + \ldots
\end{eqnarray}
The time dependent operators appearing in these expressions are
interaction picture operators, treating the full \ha (\ref{ham}) as the
bare \ha and the external coupling \ha, $H(t) = -\frac{h}{2} \sigma_z(t)
\cos{\nu t}$, as the perturbation. The $S$-matrix is expanded to the
appropriate order.  Grouping the third order terms in (\ref{S_matrix})
together, we arrive at:
\begin{eqnarray}
\<&&0|\sigma(t)I_3|0\> + \<0|I^{\dagger}_3\sigma(t)|0\> = \nonumber \\
&&\frac{-ih^3}{\hbar^3} \int_{-\infty}^{\infty}{dt_1}
\int_{-\infty}^{\infty}{dt_2} \int_{-\infty}^{\infty} 
{dt_3 \<0|\sigma(t)\sigma(t_1) \sigma(t_2) \sigma(t_3)|0\>} \nonumber \\ 
&&\times{\cos{\nu t_1}\cos{\nu t_2} \cos{\nu t_3} 
\theta(t-t_1) \theta(t_1-t_2) \theta(t_2-t_3)} \nonumber \\
&& +~ \mbox{c.c.}
\end{eqnarray} and
\begin{eqnarray}
\<&&0|I^{\dagger}_1\sigma(t)I_2|0\> + \<0|I^{\dagger}_2\sigma(t)I_1|0\>=
\nonumber \\ 
&& \frac{+ih^3}{\hbar^3} \int_{-\infty}^{\infty}{dt_1}
\int_{-\infty}^{\infty}{dt_2} \int_{-\infty}^{\infty} {dt_3
\<0|\sigma(t_1)\sigma(t) \sigma(t_2) \sigma(t_3)|0\>} \nonumber \\
&& \times{\cos{\nu t_1}\cos{\nu t_2} \cos{\nu t_3} \theta(t-t_1)
\theta(t_1-t_2) \theta(t_2-t_3)} \nonumber \\ 
&& +~ \mbox{c.c.}
\end{eqnarray}
We use the integral representation of the step function,
\begin{displaymath}
\theta(t) = \frac{1}{2\pi i} 
\int_{-\infty}^{\infty}{d\omega \frac{e^{i\omega t}}{\omega -
i\delta}}
\end{displaymath}
($\delta=0^+$) and collapse the time integrations.  From the various
products of delta functions, terms are grouped according to whether
they oscillate at $\pm 3\nu$ (third harmonic generation) or
$\pm\nu$. From the latter terms, absorptive contributions come from
the part of the signal that is $\pi/2$ out of phase with the driving
term, $-\frac{h}{2}\sigma_z \cos{\nu t}$.  

Terms that modify absorption at the original linear resonance,
$\chi^{\prime \prime} \sim P^{\prime \prime}_{n0}(\nu)$ can be grouped
as follows:
\begin{eqnarray}
\Omega_{\gamma} = && \frac{1}{4} (\frac{h}{2})^4 \nu \frac{1}{\hbar^3}
\sum_{nml}{\sigma_{0n}\sigma_{nm}\sigma_{ml}\sigma_{l0}} \nonumber \\
&& \times [P_{0n}^{\prime \prime}(-\nu)
P_{0m}^{\prime}(-2\nu)
P_{0l}^{\prime}(-\nu) \nonumber \\
&& + P_{0n}^{\prime \prime}(-\nu)
P_{0m}^{\prime}(0)
(P_{0l}^{\prime}(\nu)- P_{l0}^{\prime}(\nu))]
\end{eqnarray}
The two photon absorption terms can also be grouped as follows:
\begin{eqnarray}
\Omega_{2\gamma} = && \frac{1}{4} (\frac{h}{2})^4 \nu \frac{1}{\hbar^3}
\sum_{nml}{\sigma_{0n}\sigma_{nm}\sigma_{ml}\sigma_{l0}} \nonumber \\
&& \times P_{0m}^{\prime \prime}(-2\nu)
P_{l0}^{\prime}(\nu)
P_{n0}^{\prime}(\nu)
\end{eqnarray}

\end{document}